\documentclass{llncs}
\usepackage{fullpage}
\usepackage{makeidx}
\usepackage{algorithm}
\usepackage{algpseudocode}
\usepackage{amsmath}
\usepackage{verbatim}
\usepackage{graphicx}

\begin{document}

\title{Sorted Range Reporting Revisited\thanks{This work is a course project for CS 763 - Computational Geometry, instructed by Timothy Chan.}}

\author{Gelin Zhou}
\institute{David R. Cheriton School of Computer Science, \\ University of Waterloo, Canada. \\
\email{g5zhou@uwaterloo.ca}}

\maketitle

\bibliographystyle{splncs03}

\begin{abstract}
    We consider the two-dimensional sorted range reporting problem.
    Our data structure requires $O(n\lg\lg n)$ words of space and $O(\lg\lg n + k\lg\lg n)$ query time,
    where $k$ is the number of points in the query range.
    This data structure improves a recent result of Nekrich and Navarro~\cite{DBLP:conf/swat/NekrichN12} by a factor of $O(\lg\lg n)$ in query time,
    and matches the state of the art for unsorted range reporting~\cite{DBLP:conf/compgeom/ChanLP11}.
\end{abstract}

\section{Introduction}

The orthogonal range searching problems is well-known in the communities of computational geometry and data structures.
For these problems, we need maintain a point set $S$ in $d$-dimensional space,
such that certain functions over points in a given query rectangle $Q$ can be computed efficiently.
In this paper we study a variant of the two-dimensional orthogonal range reporting problem,
for which points in the query range are sorted in increasing order of their
$x$-coordinates~\footnote{Increasing/decreasing $x$/$y$-coordinate ordering can be easily supported via coordinate changes.}.
In addition, our data structures can work in an online fashion:
points from $S \cap Q$ are reported in increasing order of $x$-coordinates
until the query-answering procedure is terminated or all points in $S \cap Q$ are output. \\
\indent The sorted range reporting problem was proposed by Nekrich and Navarro~\cite{DBLP:conf/swat/NekrichN12}.
Let $k$ denote the number of points in the given query range.
Their linear space data structure requires $O(\lg^\epsilon n + k \lg^\epsilon n)$ query time,
and their data structure with optimal query time occupies $O(n\lg^\epsilon n)$ words of space.
These results match the state of the art for unsorted range reporting~\cite{DBLP:conf/compgeom/ChanLP11}.
However, their data structure using $O(n\lg\lg n)$ words of space requires $O(\lg^2\lg n + k \lg^2\lg n)$ query time,
which is slower than the corresponding result for unsorted range reporting~\cite{DBLP:conf/compgeom/ChanLP11} by a factor of $O(\lg\lg n)$.
In this paper we present a data structure using the same amount of space but only $O(\lg\lg n + k\lg\lg n)$ query time. \\
\indent The sorted range reporting problem is closely related to the orthogonal range successor problem
(sometimes referred to as the range next-value problem)~\cite{DBLP:conf/stacs/IliopoulosCKRW08,DBLP:journals/comgeo/YuHW11,DBLP:conf/swat/NekrichN12}.
Nekrich and Navarro's data structures~\cite{DBLP:conf/swat/NekrichN12} can be used directly to support range successor queries.
Their linear space data structure requires $O(\lg^\epsilon n)$ query time,
and their data structure achieving the optimal $O(\lg\lg n)$ query time occupies $O(n\lg^\epsilon n)$ words of space.
Our data structure requires only $O(n\lg\lg n)$ words of space to achieve the optimal query time. \\
\indent We assume that the given point set is in an $n\times n$ grid, or \emph{rank space}.
Every two points have different $x$/$y$-coordinates.
The underlying computational model throughout this work is the standard word RAM model with word size $w = \Omega(\lg n)$. \\
\indent The rest of this paper is organized as follows:
In Section~\ref{section_preliminary} we review Chan, Larsen and Patrascu's data structures~\cite{DBLP:conf/compgeom/ChanLP11} for (unsorted) range reporting
and Nekrich and Navarro's results~\cite{DBLP:conf/swat/NekrichN12} for the sorted variant.
In Section~\ref{section_main} we present our data structures for sorted range reporting and range successor queries.

\section{Preliminary}
\label{section_preliminary}

For completeness, we describe Chan et al.'s work~\cite{DBLP:conf/compgeom/ChanLP11} for unsorted range reporting
in addition to Nekrich and Navarro's work~\cite{DBLP:conf/swat/NekrichN12} for sorted range reporting.
We may modify their presentations for the sake of consistency.

\subsection{Unsorted Range Reporting}
\label{subsection_unsorted}

Chan et al.'s data structures~\cite{DBLP:conf/compgeom/ChanLP11} for range reporting
are based on the \emph{wavelet tree}~\cite{DBLP:journals/siamcomp/Chazelle88,DBLP:conf/soda/GrossiGV03}.
A conceptual range tree $\mathcal T$ is built on $[1..n]$, where $n$ is assumed to be a power of two.
Every node $v$ in $\mathcal T$ has an associated range.
Let $S(v)$ denote the set of points whose $y$-coordinates are in this range.
It is clear that $S(v)$ contains a point only for every leaf node $v$ at the bottom of $\mathcal T$.
An internal node $v$ has two children $v_l$ and $v_r$, whose associated ranges are a disjoint union of that of $v$.
Points in $S(v)$ are conceptually listed as $S(v)[1], S(v)[2], \cdots$ in increasing order of $x$-coordinates.
For each of these point, we write down a 0-bit if this point is also in $S(v_l)$, or a 1-bit otherwise.
These bits are concatenated and maintained as a bit vector,
such that rank/select operations can be performed in constant time~\cite{DBLP:conf/soda/ClarkM96}. \\
\indent To achieve efficient query time,
Chan et al.~\cite{DBLP:conf/compgeom/ChanLP11} formulated the following two operations as the \emph{ball-inheritance problem}:
Given a node $v$ in $\mathcal T$ and a range $[a..b]$ on the $x$-axis,
$noderange(v, a, b)$ returns the range $[a_v..b_v]$ such that
$S(v)[a_v]$ is the first one whose $x$-coordinate is $\geq a$
and $S(v)[b_v]$ is the last one whose $x$-coordinate is $\leq b$.
Given a node $v$ in $\mathcal T$ and an index $1 \leq i \leq |S(v)|$,
$point(v, i)$ returns the coordinates of $S(v)[i]$.
The following lemma addresses their results.
\begin{lemma}[\cite{DBLP:journals/siamcomp/Chazelle88,DBLP:conf/compgeom/ChanLP11}]
    \label{lemma_ball_inheritance}
    Using $O(nf(n))$ words of space,
    one can support operations $noderange(v, a, b)$ and $point(v, i)$ with $O(g(n) + \lg\lg n)$ and $O(g(n))$ query time, respectively,
    where
    \begin{enumerate}
        \item $f(n) = O(1)$ and $g(n) = O(\lg^\epsilon n)$;
        \item $f(n) = O(\lg\lg n)$ and $g(n) = O(\lg\lg n)$;
        \item $f(n) = O(\lg^\epsilon n)$ and $g(n) = O(1)$.
    \end{enumerate}
\end{lemma}

\indent A range reporting query $Q = [a..b]\times[c..d]$ over the point set $S$ can be answered as follows.
First we find node $v$ in $\mathcal T$, which is the lowest common ancestor of the leaf nodes that correspond to $c$ and $d$.
Let $v_l$ and $v_r$ denote the children of $v$.
It is clear that the associated range of $v$ contains $[c..d]$,
and the associated ranges of $v_l$ and $v_r$ both intersect $[c..d]$.
Therefore, $S \cap Q$ is decomposed into $S(v_l) \cap ([a..b]\times [c..\infty])$ and $S(v_r) \cap ([a..b]\times [-\infty..d])$.
We consider how to support $S(v_r) \cap ([a..b]\times [-\infty..d])$ only.
We compute $[a_{v_r}..b_{v_r}]$ using $noderange(v_r, a, b)$.
Thus, we need only report all the points in $S(v_r)[a_{v_r}..b_{v_r}]$ whose $y$-coordinates $\leq d$.
To perform this step efficiently, an index for range minimum queries over $S(v_r)$ is built.
This index returns the position of the point with the smallest $y$-coordinate in $S(v_r)[i..j]$ using constant time
and $2|S(v_r)| + o(|S(v_r)|)$ bits of space~\cite{DBLP:journals/siamcomp/FischerH11}.
The following paragraph shows how to report $k$ points in $kg(n)$ time: \\
\indent Initially, we set $[i..j] = [a_{v_r}..b_{v_r}]$.
We query $[i..j]$ on the index for range minimum queries, and $l$ is returned.
Using Lemma~\ref{lemma_ball_inheritance}, we verify if the $y$-coordinate of $S(v_r)[l]$ is no greater than $d$.
We terminate if the $y$-coordinate is greater.
Otherwise, we report $S(v_r)[l]$ and recurse on $[i..l - 1]$ and $[l + 1..j]$. \\
\indent It is noteworthy that the first point returned by this algorithm is the lowest point in $S(v_r) \cap Q$.
On the other hand, the first point returned by the other 3-sided query is the highest point in $S(v_l) \cap Q$.
We cannot simply obtain the lowest or highest point in $S \cap Q$. \\
\indent The following lemma summarizes the discussions in this section.
\begin{lemma}[\cite{DBLP:conf/compgeom/ChanLP11}]
    \label{lemma_unsorted_range_reporting}
    Using $O(nf(n))$ words of space,
    one can support two-dimensional range reporting queries with $O(\lg\lg n + g(n) + k g(n))$ query time,
    where
    \begin{enumerate}
        \item $f(n) = O(1)$ and $g(n) = O(\lg^\epsilon n)$;
        \item $f(n) = O(\lg\lg n)$ and $g(n) = O(\lg\lg n)$;
        \item $f(n) = O(\lg^\epsilon n)$ and $g(n) = O(1)$.
    \end{enumerate}
\end{lemma}

\subsection{Suboptimal Range Successor}

\indent Nekrich and Navarro~\cite{DBLP:conf/swat/NekrichN12} showed that
Chan et al.'s data structures~\cite{DBLP:conf/compgeom/ChanLP11} can support range successor queries after certain modifications.
For simplicity, we swap $x$-/$y$-coordinates and return the lowest point (i.e., the one with the smallest $y$-coordinate)
in the query range $Q = [a..b]\times[c..d]$.
Let $\pi$ denote the path from the root of the conceptual range tree $\mathcal T$ to the leaf that corresponds to $c$.
The basic idea is to find the lowest node $v_f$ on $\pi$ such that $S(v_f) \cap Q \neq \phi$.
Let $v$ denote the lowest common ancestor of the leaf nodes that correspond to $c$ and $d$.
It is clear that, for every node $u$ on $\pi$ that is deeper than $v$, its associated range contains $c$ but not $d$.
It implies that $S(u) \cap Q = S(u) \cap ([a..b]\times[c..\infty])$.
One can determine if $S(u) \cap Q \neq \phi$ by determining if the 3-sided query contains any point.
Therefore $v_f$ can be computed using binary search on $\pi$.
This requires to compute $O(\lg\lg n)$ 3-sided emptiness queries. \\
\indent If $v_f$ is a leaf node, then the only point in $S(v_f)$ is the answer.
If $v_f$ is an internal node, the child of $v_f$ on $\pi$ must be the left child.
Otherwise the associated range of the left child would not intersect $[c..d]$,
and the assumption on $v_f$ would be contracted.
Since the left child of $v_f$ does not correspond to any point in the query range $Q$,
the right child must correspond to at least a point in $Q$.
Using the algorithm described in the previous section, we can find such a point using range minimum queries.
The first point returned is by chance the lowest one. \\
\indent The following lemma summarizes the above discussions.
\begin{lemma}[\cite{DBLP:conf/swat/NekrichN12}]
    Using $O(nf(n))$ words of space,
    one can support two-dimensional range successor queries with $O(g(n)\lg\lg n)$ query time,
    where
    \begin{enumerate}
        \item $f(n) = O(1)$ and $g(n) = O(\lg^\epsilon n)$;
        \item $f(n) = O(\lg\lg n)$ and $g(n) = O(\lg\lg n)$.
    \end{enumerate}
\end{lemma}

One can support sorted range reporting queries using range successor queries.
Suppose the lowest point in $Q = [a..b]\times[c..d]$ has $y$-coordinate $p$.
We can find the second lowest point in $Q$ by querying $Q' = [a..b]\times[p+1..d]$.
The procedure is repeated until the query range contains no point.
Thus we have the following lemma:
\begin{lemma}[\cite{DBLP:conf/swat/NekrichN12}]
    Using $O(nf(n))$ words of space,
    one can support two-dimensional sorted range reporting queries with $O(\lg\lg n(g(n) + k g(n)))$ query time,
    where
    \begin{enumerate}
        \item $f(n) = O(1)$ and $g(n) = O(\lg^\epsilon n)$;
        \item $f(n) = O(\lg\lg n)$ and $g(n) = O(\lg\lg n)$.
    \end{enumerate}
\end{lemma}

\section{Optimal Range Successor Queries in Less Space}
\label{section_main}

We present the main result in this paper:
a data structure for range successor queries using $O(n\lg\lg n)$ words of space and $O(\lg\lg n)$ query time.
This data structure is obtained by modifying Nekrich and Navarro's third data structure~\cite{DBLP:conf/swat/NekrichN12} for sorted range reporting.
We consider 3-sided range successor queries,
for which the leftmost point in $S \cap ([a..b]\times[-\infty..d])$ is returned.
A preliminary result of Nekrich and Navarro is addressed in the following lemma.
\begin{lemma}[Lemma 5 in \cite{DBLP:conf/swat/NekrichN12}]
    \label{lemma_three_sided}
    Given a set of $n$ points, one can support 3-sided range successor queries
    using $O(n\lg^3 n)$ bits of space and $O(\lg\lg n)$ query time.
\end{lemma}

\indent Now we describe our data structures.
We first build the same data structures as the second variant of Lemma~\ref{lemma_unsorted_range_reporting}.
Let $Q' = [a..b]\times[-\infty..d]$ denote a 3-sided query range.
We further construct auxiliary data structures on every $S(v)$ such that
the leftmost point in $S(v) \cap Q'$ can be returned in $O(\lg\lg n)$ time, using $O(|S(v)|\lg\lg n)$ bits of additional space. \\
\indent As we have mentioned, points in $S(v)$ are conceptually listed in increasing order of $x$-coordinates.
These points are divided into blocks $B_1(v), B_2(v), \cdots$ of size $\lceil \lg^3 n\rceil$ (the last block may contain less).
$D(v)$ stores the lowest point in each block explicitly,
and is maintained using Lemma~\ref{lemma_three_sided} such that 3-sided range successor queries on $D(v)$ can be supported in $O(\lg\lg n)$ time.
This auxiliary data structure occupies $O(|D(v)|\lg^3 |D(v)|) = O(|S(v)|)$ bits of space. \\
\indent For some constant $0 < \epsilon < 1$, points in every block $B_i(v)$ are further divided into sub-blocks
$SB_{i,1}(v), SB_{i,2}(v), \cdots$ of size $\lceil \lg^\epsilon n \rceil$ (the last sub-block may contain less).
In addition, their $x$/$y$-coordinates are rewritten as the $x$/$y$-ranks within this block.
A rank can be represented in $O(\lg\lg n)$ bits, and all the ranks require $O(|S(v)|\lg\lg n)$ bits over all blocks.
$E_i(v)$ stores the lowest point in every sub-block explicitly,
and is also maintained using Lemma~\ref{lemma_three_sided} to support 3-sided range successor queries on $E_i(v)$ in $O(\lg\lg\lg n)$ time.
This auxiliary data structure requires only $O(|E_i(v)|\lg^3 |E_i(v)|) = O(|B_i(v)|\lg^3\lg n / \lceil \lg^\epsilon n \rceil) = o(|B_i(v)|)$
bits of additional space.
Thus the space cost of all $E_i(v)$'s is $o(|S(v)|)$ bits. \\
\indent Now we show how to answer the 3-sided range successor query $Q' = [a..b]\times[-\infty..d]$ over $S(v)$.
First we compute $[a_v..b_v]$ using $noderange(v, a, b)$, which requires $O(\lg\lg n)$ time.
Let $B_i(v)$ and $B_j(v)$ be the blocks that contains $a_v$ and $b_v$, respectively.
Thus $[a_v..b_v]$ spans over blocks $B_i(v), \cdots, B_j(v)$.
We first attempt to find the leftmost point in $B_i(v) \cap Q'$ (we will show how to do it later).
If such a point exists, our algorithm terminates and the point is returned.
Otherwise, we query $D(v)$ to find the leftmost block among $B_{i+1}(v), \cdots, B_{j-1}(v)$ that intersect $Q'$.
Let $B_l(v)$ denote the block.
We query $B_l(v) \cap Q'$ and returns the result.
If such $l$ does not exist, we query $B_j(v) \cap Q'$. \\
\indent The remaining issue is to find the leftmost point in $B_i(v)$ that is contained in a given 3-sided range $Q'$ efficiently.
We first compute the ranks of $a$, $b$ and $d$ within this block.
Let them be $a'$, $b'$ and $d'$, respectively.
$a'$ and $b'$ can be computed directly from $a_v$ and $b_v$.
The computation of $d'$ requires succinct indices for predecessor search~\cite{DBLP:conf/stacs/GrossiORR09}, using $O(\lg\lg n)$ time.
Similar to the procedure described in the previous paragraph,
we find the leftmost point in $B_i(v) \cap ([a'..b']\times[-\infty..d'])$ in $O(\lg\lg n)$ time.
The only difference is that we find the leftmost point within a sub-block using table-lookup.
This can be done using constant time and a global lookup table of $o(n)$ bits of additional space,
since there are only $2^{\lceil \lg^\epsilon n \rceil \times O(\lg\lg n)} = O(n^{1 - \delta})$ different sub-blocks,
for some positive constant $\delta$. \\
\indent Summarizing the discussion above, we can find the leftmost point in $S(v) \cap Q'$ in $O(\lg\lg n)$ time.
We also construct auxiliary data structures on every $S(v)$ for 3-sided range successor queries
of the form $[a..b]\times[c..\infty]$.
Combining both parts and following the approach in Section~\ref{subsection_unsorted},
we can find the leftmost point in a 4-sided query range in $O(\lg\lg n)$ time.
\begin{theorem}
    Using $O(n\lg\lg n)$ words of space, one can support two-dimensional range successor queries in $O(\lg\lg n)$ time.
    Repeatedly using this data structure, one can support two-dimensional sorted range reporting queries in $O(\lg\lg n + k\lg\lg n)$ time,
    where $k$ is the size of output.
\end{theorem}

\bibliography{ref}

\end{document}